\documentclass[preprint,showkeys,showpacs]{revtex4}

\usepackage{amsmath}
\usepackage{amssymb}
\usepackage{graphicx}

\begin{document}

\def\bea{\begin{eqnarray}}
\def\eea{\end{eqnarray}}
\floatsep 30pt
\intextsep 36pt

\title{Exact solution to the Schr\" odinger's equation with pseudo-Gaussian potential.}

\author{Felix Iacob}
\email{felix@physics.uvt.ro, felix.iacob@gmail.com} 
\affiliation{{\small \it  West University of Timi\c soara,\\
  300223 V. P\^ arvan 4, Timi\c soara, Romania. }}

\author{Lute Marina}
\email{  marina.lute@upt.ro}
\affiliation{{\small \it  Politehnica University of Timi\c soara,\\
  300223 Traian Lalescu 2, Timi\c soara Romania.\\}}

\begin{abstract}
\noindent We consider the radial Schr\" odinger equation with the pseudo-Gaussian potential. By making an ansatz to the solution of the  eigenvalue equation for the associate Hamiltonian, we arrive at the general exact eigenfunction. 
The values of energy levels for the bound states are calculated along with their corresponding normalized wave-functions.
The case of positive energy levels, known as meta-stable states, is also discussed and the magnitude of transmission coefficient through the potential barrier is evaluated.
\end{abstract}



\maketitle

\section{Introduction}\label{intro}
There are nearly a hundred years since the Schr\" odinger's famous equation was published \cite{Sch} and
a large number of attempts have been made towards the exact and approximate calculation of its eigenvalues and eigenfunctions employing analytic, semi-numerical or purely computational methods.
Exactly solvable Schr\" odinger equations (SE) play an important role in quantum physics and their solutions have a wide applicability into atomic and molecular physics, solid-state physics and chemistry. 
The successful examples are both the harmonic oscillator and the hydrogen atom. 
Other physical systems, which admit exact solutions of SE, are described by potentials known as:
the Morse potential \cite{M}, the Eckart potential \cite{Eck}, the Rosen-Morse potential \cite{RM}, the trigonometric and hyperbolic P\" oschl-Teller potentials \cite{PT},  the Manning-Rosen potential \cite{MR}, the Woods-Saxon potential \cite{WS} and  the Scarf potential \cite{Sca}. In these cases the SE takes the form of a hypergeometric equation, an exception being made by the Morse potential, which makes SE to become a confluent equation. 
There are a few known methods to solve exactly the SE. One method is to transform it into a known ordinary differential equation (ODE) that admits solutions in the special functions class, this includes Hermite, Legendre, Bessel, Whittaker, Heun, Kummer functions and other (confluent) hypergeometric functions. The Nikiforov-Uvarov method was proposed and applied to reduce the second order differential equation to an hypergeometric type equation by an appropriate coordinate transformation. Another method is the factorization of Hamiltonian, or generally the so  called Supersymmetric Quantum Mechanics (SUSY-QM) method. One can found many references on these methods, so we do not insist on the subject. There is a method where an ansatz is made for finding the exact solutions of a certain differential equation. This was first made, by  Hans Bethe in 1931, to find the exact eigenvalues and eigenvectors of the one-dimensional anti-ferromagnetic Heisenberg model Hamiltonian \cite{Bethe}. Since then the method has been developed and extended to other physical systems.  We will mention Ogata and Shiba \cite{Ba}, where the momentum distribution function and spin-correlation function are determined for the Hubbard model with various electron densities by an Bethe-ansatz on wave-function, and Kaushal's work \cite{Kaushal_other} 
where the solutions of SE are investigated for a variety of potentials as a further application of an already suggested ansatz in \cite{Kaushal}.  The D-dimensional SE was exactly solved for some anharmonic, pseudo-harmonic and Kratzer potentials making an ansatz to the wave-functions \cite{dong_ans}. The ``ansatz method'' was also applied to inverse-power potentials \cite{dong_inv} and to radial SE for some physical potentials \cite{Ikhadir}.

In this work we obtain an exact analytic solution of radial Schr\" odinger equation for the pseudo-Gaussian potential by making an ansatz to the wave-functions. The related physical model, named pseudo-Gaussian oscillator (PGO), was introduced in \cite{GF,F} and the energetic levels were determined using the generating functional method (GFM).
PGO are a family of potentials with finite number of energy eigenstates where the most important feature is that, the PGO potentials have the harmonic oscillator (HO) properties near zero together with a Gaussian asymptotic behavior. We would like to outline that PGO may be considered as an option to describe physical systems with an oscillating behavior. In order to be able to solve the SE, we had to notice that the pseudo-Gaussian potential can be written as an infinite power series and consider it as a polynomial of infinite degree. 
Rainville, in 1936, introduced the ansatz method for the polynomial solutions of Riccati equations \cite{Rainville}. In a standard manner Riccati equation can be reduced to a second-order linear differential equation.
Solutions of SE with polynomial potentials up to $6-th$  degree, known as doubly anharmonic or sextic potential, \cite{Kaushal,dong,Nasser6} and up to $10-th$ degree \cite{octic,dong10,Nasser}, known as decatic potential, have been reported. To complete this work, for pseudo-Gaussian potential, we had to avoid some difficulties  encountered within the infinite case. We relied on the fact that pseudo-Gaussian potential has Gaussian asymptotic behavior, thus the corresponding power series will not diverge and get solutions in the set of square-integrable function, $L^2({\mathbb R})$.

We divided this paper into two parts. The first one (section \ref{gs}) presents the pseudo-Gaussian model, introducing some of its properties and  gives the general solution of SE for this. The second one (section \ref{spc}) is devoted to applications and particular cases. The energy levels are calculated and modifications of the energy levels, at change of potential parameters, are shown. 
This section, also, includes the graphs of first eigenfunctions, for each quantum number.

\section{The physical model. General solutions.}\label{gs}

\begin{figure}[h!]
  \centering
    \includegraphics[scale=0.3]{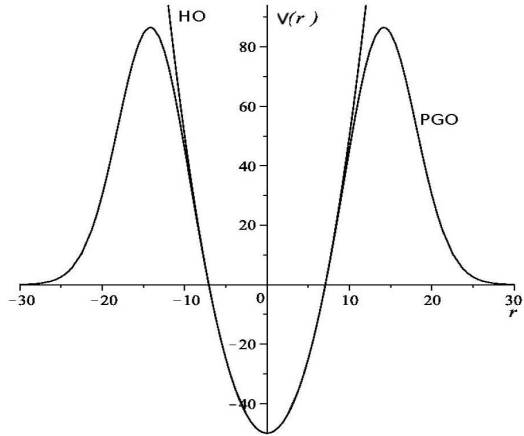}
  \caption{The pseudo-Gaussian oscillator potential graph ($s=3$) compared with harmonic oscillator potential one.}
\label{PGO_pot}
\end{figure}

Let us consider the radial part of the three-dimensional Schr\" odinger's time-independent equation, $\mathbf H \psi = E\psi$, 
the square-integrable complex functions  $\psi$ of real variable are called eigenfunctions and the numbers $E$ are called eigenvalues of the energy. 
The Hamiltonian operator $\mathbf H$, acting on the space of eigenfunctions, given in atomic units,
$
\mathbf H = -\frac{1}{2}\Delta + V(r),
$
introduces the central real valued potential $V(r)$ on Euclidean real space with spherical coordinates as:
\begin{equation}\label{pot}
V_{\lambda,\mu}^s(r)=\left(\lambda+\sum_{k=1}^s C_k
r^{2k}\right)\exp(-\mu r^2)\,,
\end{equation}
having the coefficients $C_k$, \cite{F}:
\begin{equation}\label{Ck}
C_k=\frac{(\lambda+k)\mu^{k}}{k!}\,.
\end{equation}
The properties of this model are completely determined by the
dimensionless parameters $\lambda\in {\mathbb R}$, $\mu\in {\mathbb R_+}$ and the positive
integer $s=1,2,...$, named the {\em order} of PGO.
We note that the genuine Gaussian potential corresponding to the
order $s=0$ is not included in this family.
The potentials defined by the eqs. (\ref{pot}) and (\ref{Ck}) have the
remarkable property to approach to the HO potential when $r\to 0$ and together with Gaussian asymptotic behavior, i.e. $\lim_{r\to\infty} V_{\lambda,\mu}^s(r)=0$.
We also have to notice that, for each order, $s$, the Taylor
expansion of these potentials does not have terms proportional with $r^4,r^6,...,r^{2s}$.
\begin{equation}\label{Tay}
V_{\lambda,\mu}^s(r)=\lambda + \mu r^2 + O(r^{2s+2})\,
\end{equation}
In figure (\ref{PGO_pot}) it is shown the graph of both PGO and HO, one can see  their similar shape in a vicinity of origin and 	
the Gaussian asymptotic behavior of PGO beside HO, which goes to infinity.

Taking into consideration that  $l(l + 1)$ represents the eigenvalue of the square of the angular-momentum operator $\mathbf L^2 $, the radial part of the Schr\"{o}dinger equation for the stationary states can be written as: 
\begin{equation}\label{Swd}
\left[\frac{d^2}{dr^2}+\frac{(2)}{r}\frac{d}{dr}
+(E-V(r))-\frac{l(l+1)}{r^2}\right]\psi(r)=0, 
\end{equation}
with $\psi(r)$ the radial wave-function. Considering the potential (\ref{pot}) we can study the radial three dimensional Schr\" odinger eigenvalue problem. 
We can  eliminate the first derivative by setting 
\begin{equation}\label{solpsi}
 \psi(r)\equiv r^{-1}R(r)
\end{equation}
and (\ref{Swd}) becomes: 
\begin{equation}\label{Swod}
\left[\frac{d^2}{dr^2}+(E-V(r))-\frac{l(l+1)}{r^2}\right]R(r)=0.
\end{equation}

We can calculate the series (\ref{Tay}) of potential (\ref{pot}) with coefficients (\ref{Ck}), and after some algebra we get the form:
\begin{equation}\label{pot_t}
V_{\lambda,\mu}^s(r) = \lambda + \mu r^2 + \sum_{k=s+1}\hat C_k r^{2k},
\end{equation}
with coefficients $\hat C_k$ coming from the Taylor expansion and having the following form: 
\begin{equation}\label{hatC}
\hat C_k \propto (-1)^{s+k} \left(\frac{1}{(k-1)!}+\frac{\lambda}{k!}\right) (k-s)^\frac{k}{2}.
\end{equation}
The potential (\ref{pot}) with terms grouped as in (\ref{pot_t}), allows us to recognize that it is basically made up by a term representing the HO potential $V_{HO}=\lambda + r^2$, with $\lambda$ an arbitrary energy level and an additional term, which is actually a power series, $V_{int}= \sum_{k=s+1}\hat C_n r^{2k}$.
At this point one can easily investigate the solutions of SE by a perturbative method. 
We will let the perturbation theory approach as an later work and follow the standard techniques 
 by applying the SAP method (Simplify, Asymptote, Power Series) to solve the second-ODE given by (\ref{Swod},\ref{pot_t}). However, the method used to HO cannot be applied in this case due to terms proportional with $r^{2k}$, ($k>s+1$) coming from $V_{int}$. 

Our attention was directed to an ansatz that will encompass these terms, having exponents of $r$ higher than two, by using a suitable polynomial at the exponent of the Euler's function. The idea is that in the power series method we are equating the $(n+1)-th$ coefficient to zero to find a relation that will stop the power series to diverge. To do that, in our case, we need to cancel the terms with exponents strictly greater than two in the SE, introduced by potential (\ref{pot_t}). 
This requirement will be accomplished by our ansatz upon the solutions, with the consequence that this introduces more equations to the condition, which will stop the power series to diverge, as will be shown.
Following \cite{Kaushal,dong}, we find that this suitable ansatz for the radial wave-functions can be written as:
\begin{equation}\label{ans}
R(r)=\exp[p(r)]\sum\limits_{n=0}a_{n}r^{2n+\tau},
\end{equation}
the polynomial $p(r)$ has the form:
\begin{equation}\label{pol}
p(r) = \sum\limits_{n=1}^{s+1} \frac{1}{2n} \alpha_{2n}r^{2n},
\end{equation}
where $\alpha_{2n}$ are some parameters that will be determined.
By inserting (\ref{ans},\ref{pol}) into the radial SE we find the following relations after equating the coefficient of $r^{2n+\tau}$ to zero:
\begin{equation}\label{equ0}
\sum\limits_{j=s+1} \left (A_j\right)_{n}a_{n+j}
+\sum\limits_{j=1}^{s} \left (S_j\right)_{n}a_{n-s+j}
+\left (S_0\right)_{n}a_{n}
+B_{n}a_{n+1}+C_{n}a_{n+2}=0
\end{equation}
where the calculated expressions of involved terms are:

\begin{equation}\label{Auri}
\left(A_j\right)_n = 2\alpha_{i}\alpha_{2s}  + \sum_{i=j-s+2}^{n-1}\alpha_{2i}\alpha_k \,\delta(2j+4-2i-k)-\hat C_{j+1},
\;\;\; j \in \{s+1,s+2,\,\ldots\}
\end{equation}
\begin{equation}\label{Suri}
\left(S_j\right)_n = 2\alpha_{2j+4}(\tau+2n) + (2j+3)\alpha_{2j+4} + \sum_{i=1}^{s-1}\alpha_{2i}\alpha_k \,\delta(2j+4-2i-k),
\;\;\; j \in \{1,\, s\}
\end{equation}
\begin{subequations}
\begin{align}\label{S0}
\left(S_0\right)_n &= 2\alpha_2^2 + (3+2\tau+4n)\alpha_4-\mu^2,\\\label{B}
B_n&=(1+2\tau+2n)\alpha_2 +E-\lambda,\\\label{C}
C_n&=(\tau+4n)(-1+\tau+2n)) - l(l+1).
\end{align}
\end{subequations}
These notations are a generalization of those used in (\cite{dong}) and the well known Kronecker symbol, ($\delta(i)=1,\; i=0$ and $\delta(i)=0,\; i\neq 0$).

An important observation is that the Eqs. (\ref{Auri}) do not depend explicitly on $n$, just the number of terms involved in these relations are proportional with $n$. The parameters $\alpha_{2n}$ can be determined uniquely from these equations and this ensures the cancellation of higher order terms in SE coming from $V_{int}$. 

Let us see the technical steps followed to determine the wave-function and its associated energy levels. Considering the first non-vanishing coefficient $a_{0}$ in Eq. (\ref{equ0}), we have to impose the condition $C_0=0$, so the expression (\ref{C}), with $n=0$, gives the values for $\tau$. In order to have well behaved wave-function solutions we choose $\tau=l(l+1)$.
The values $\alpha_i$, coefficients of polynomial $p(r)$, are obtained solving the system made up from Eqs. (\ref{Auri}). 
The condition of having non-trivial solution for the system of remaining equations is that $(S_j)_n$, $B_n$, $C_n$ satisfy the determinant relation
\begin{equation}\label{deteq}
\det\left|
\begin{array}{llllllll}
B_{0} & C_{1}&0&0&\cdots&\cdots&0&0\\
S_{0}& B_{1}& C_{2}& 0 &\cdots&\cdots&0&0\\
(S_{1})_0& (S_{0})_1&B_{2}& C_{3}& \cdots&\cdots&0&0 \\
\vdots&\vdots&\vdots&\ddots&\vdots&\vdots&\vdots&\vdots\\
(S_{s})_0& (S_{s-1})_1&(S_{s-2})_2& (S_{s-2})_3& \cdots&\cdots&0&0 \\
0&(S_{s})_1& (S_{s-1})_2&(S_{s-2})_3&  \cdots&\cdots&0&0 \\
0&0&(S_{s})_2&(S_{s-1})_3&  \cdots&\cdots&B_{n-1}&C_{n}\\
0&0&0&(S_{s})_3&  \cdots&\cdots&(S_0)_{n-1}&B_{n}\\
\end{array}\right|=0,
\end{equation}
which is the energy levels quantification condition.
That being said, with the coefficients $\alpha_i$ for the polynomial (\ref{pol}) and the conditions (\ref{deteq}) for a convergent solution of SE (\ref{Swod}), the corresponding eigenfunctions, by virtue of relation (\ref{solpsi}), can be now written as:
\begin{equation}\label{sol}
\psi_n(r)=N_n\, r^{l(l+1)-1} \exp[p_n(r)].
\end{equation}
$N_n$ denotes a polynomial of degree n,  whose coefficients $a_i$, introduced by rel. (\ref{ans}), are determined from the normalization condition, this means the eigenfunctions obey the condition of square-integrable functions,
\begin{equation}\label{normcond}
\int_0^\infty \left|\psi_n(r)\right|^2 r^2dr = 1.
\end{equation}
In this way, we can generate a class of exact solutions through setting $n= 0,1,2,\ldots$.

\section{Discussions and some particular cases.}\label{spc}

Before we proceed with the analysis of individual cases, we must make some remarks on Eqs. (\ref{pot_t}) and (\ref{equ0}), in order to be able to give the numeric values for the energy levels and to find their corresponding eigenfunctions. These relations have to be understood as a limit process when $N\to\infty$. $N$ is a natural positive number that measures the closeness of the $\left(V_{int}\right)_N = \sum_{k=s+1}^N\hat C_k r^{2k}$ to the genuine potential $V_{int}$, introduced the in previous section. The higher is $N$, the closer is $\left(V_{int}\right)_N$ to $V_{int}$. This is just like when we want to give Euler's constant a numeric value. In order to improve its accuracy, we must always add a new term to its power series definition. This concept  is also promoted, in the theory of differential equations, as the case of quasi-exactly solvable systems \cite{QSS}.

\begin{figure}[h]
  \centering
    \includegraphics[scale=0.4]{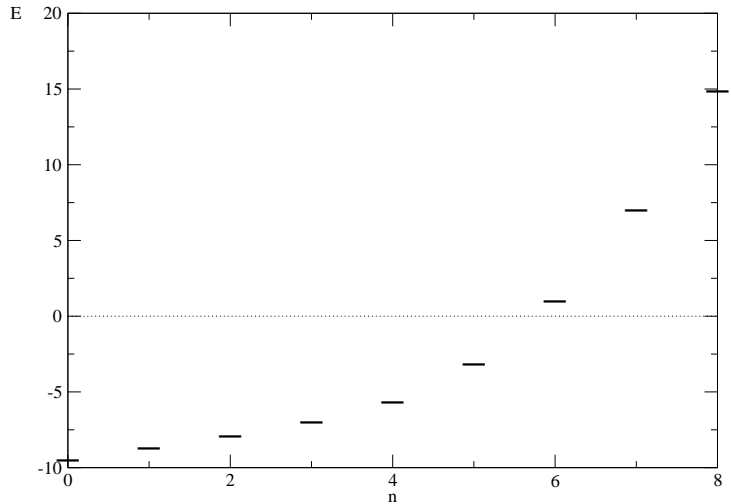}
  \caption{The energy levels for pseudo-Gaussian oscillator, $N=7$, $\lambda=-5.6$, $\mu=0.2$ and $l=0$.}
\label{en7}
\end{figure}

Having this remark in mind, Eqs. (\ref{pot_t}) and (\ref{equ0}) can be rewritten:
\begin{equation}\label{pot_t_N}
\left (V_{\lambda,\mu}^s(r)\right )_N = \lambda + \mu r^2 + \sum_{k=s+1}^N \hat C_k r^{2k}
\end{equation}
\begin{equation}\label{equ0_N}
\sum\limits_{j=s}^{N-1} \left (A_j\right)_{n}a_{n+j}
+\sum\limits_{j=1}^{s} \left (S_j\right)_{n}a_{n-s+j}
+\left (S_0\right)_{n}a_{n}
+B_{n}a_{n+1}+C_{n}a_{n+2}=0
\end{equation}
Let us present an algebraic evaluation with seven terms, $N=7$, s=3. The polynomial, (\ref{pol}), takes the form:
\begin{equation}\nonumber
p(r) = \frac{1}{2} \alpha_2 r^2+  \frac{1}{4}\alpha_4r^4 + \frac{1}{6}\alpha_6r^6 + \frac{1}{8}\alpha_8 r^8,
\end{equation}        
and the Eqs. (\ref{Auri}), wherefrom coefficients $\alpha_i$ can be determined, are written: 
\begin{subequations}\nonumber
\begin{align}
     A_3 &= 2 \alpha_8 \alpha_2 + 2 \alpha_4 \alpha_6 - \hat C_4\\
     A_4 &=  2 \alpha_4 \alpha_8 + \alpha_6^2  - \hat C_5\\
     A_5 &=  2 \alpha_6 \alpha_8 -  \hat C_6\\  
     A_6 &=  \alpha_8^2  -  \hat C_7
\end{align}
\end{subequations}
having the associated matrix upper triangular. 
As far as our calculation goes this upper triangular form is preserved for any value of $N$.  
We can find the energies from the condition (\ref{deteq}), this is: 
\begin{equation}\label{det73}
\det\left|
\begin{array}{lllllll}
B_{0} & C_{1}&0&0&0&0&0\\
(S_{0})_0& B_{1}& C_{2}& 0 &0&0&0\\
(S_{1})_0& (S_{0})_1&B_{2}& C_{3}& 0& 0&0 \\
(S_{2})_0& (S_{1})_1&(S_{0})_2& B_3& C_{4}&0&0 \\
0&(S_{2})_1& (S_{1})_2&(S_{0})_3&  B_4& C_{5}&0 \\
0&0&(S_{2})_2&(S_{1})_3& (S_{0})_3&B_{5}&C_{6}\\
0&0&0&(S_{2})_3& (S_{1})_4&(S_0)_{5}&B_{6}\\
\end{array}\right|=0.
\end{equation}
The calculated values of energy for the pseudo-Gaussian oscillator system are presented in figure (\ref{en7}).
We have to admit that an analytic relation cannot be established, because while solving (\ref{det73}), we had to face with polynomial equations of degree higher than five. It is known that Abel's impossibility theorem states that there is no general algebraic solution, formulate in radicals, to polynomial equations of fifth or higher degree. 

\begin{figure}[h]
  \centering
    \includegraphics[scale=0.4]{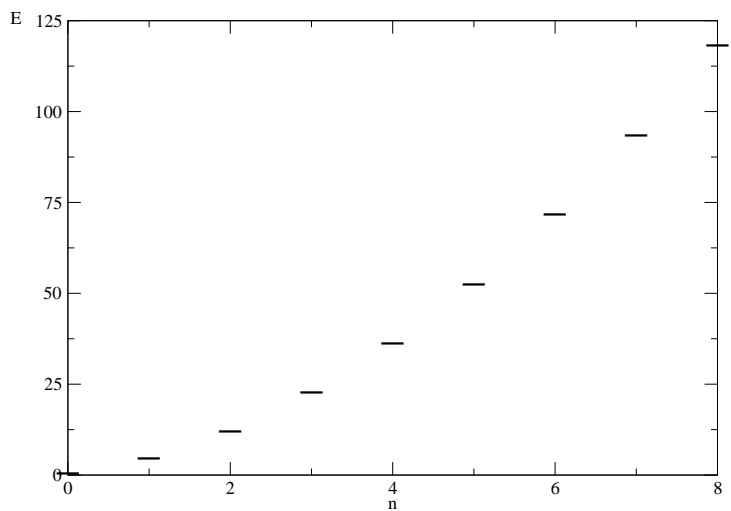}
  \caption{The energy levels for pseudo-Gaussian oscillator, $N=11$, $\lambda=0$, $\mu=0.2$ and $l=0$.}
\label{en11lmbd0}
\end{figure}

Hereinafter, considering the reference energy $\lambda=0$, we get the energies for the pseudo-Gaussian oscillator presented in figure (\ref{en11lmbd0}).
This result is in agreement with that one obtained numerically by generating functional method \cite{GF}.

\begin{figure}[h]
  \centering
    \includegraphics[width=9cm,height=7.9cm]{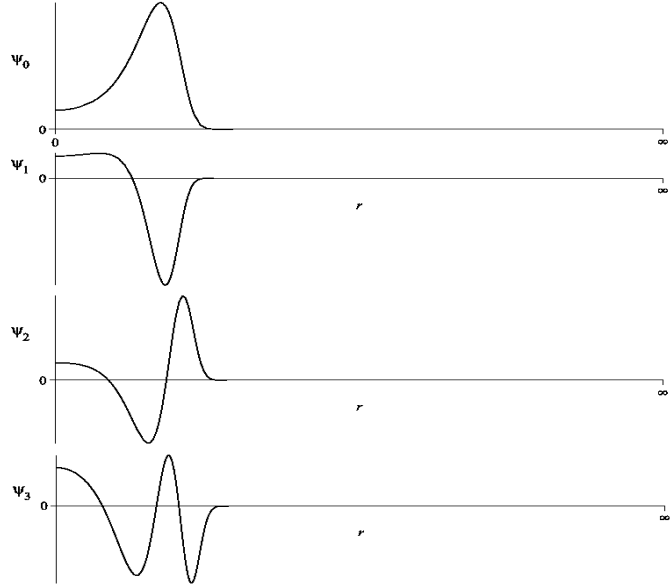}
  \caption{The first four eigenfunctions for pseudo-Gaussian oscillator, $N=7$, $\lambda=-5.6$, $\mu=0.2$ and $l=0$.}
\label{eigfunc}
\end{figure}

Let us see the analytic expression of ground wave eigenfunction:
\begin{equation}\label{sol7_0}
\psi_0(r)=a_0\, r^{-1} \exp[0.638 r^2 - 0.039 r^4 + 0.0034 r^6 - 0.0029 r^8],
\end{equation}
where $a_0$ has the calculated value from condition of square-integrable functions (\ref{normcond}), $\frac{1}{a_0}=5.081935531$. 
In figure (\ref{eigfunc}) we represent the graph of wave eigenfunctions for $n\in \{0,1,2,3\}$.
The coefficients $a_i$ of normalization polynomial $N_n$ are calculated step by step such that relation (\ref{normcond}) holds.

\begin{figure}[h]
  \centering
    \includegraphics[width=9cm,height=7.0cm]{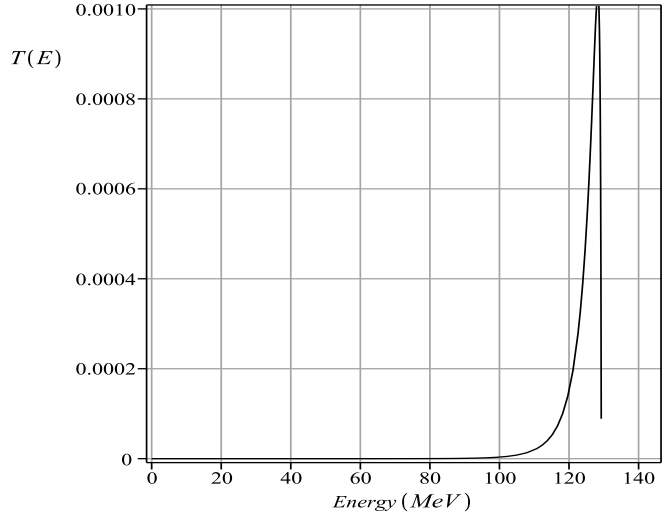}
  \caption{The transmission coefficient for the highest energy meta-stable state.}
\label{tr}
\end{figure}

The shape of PGO potential, see fig. (\ref{PGO_pot}), appears like a well surrounded by barriers. Thus, this potential admits  eigenstates with positive energies, known as meta-stable states or resonances. 
A particle that is bound by some attractive force is able to escape even though it lacks the energy to overcome the attractive force. Classical physics predicts that such behavior is impossible. However, the fuzziness of nature at the sub-atomic scale, which is an inherent part of quantum mechanics, implies that we cannot know precisely the trajectory of α particle, this uncertainty means that
the particle has a small, but non-zero probability of suddenly finding itself outside. We say it has tunneled through a potential energy barrier created by the attractive force. 
The transmission coefficient for a particle tunneling through a potential barrier is:
\begin{equation}\label{tc}
T = \exp\left ( -2\int_{r_1}^{r_2} dr \sqrt{\frac{2m}{\hbar^2} (V(r)-E)}\right ).
\end{equation}
The effective calculation will be made with the help of Gamow factor, by chopping the PGO potential into infinitesimal potential steps with constant values of length, let say, $L$. 
\begin{figure}[h]
  \centering
    \includegraphics[width=9cm,height=7.0cm]{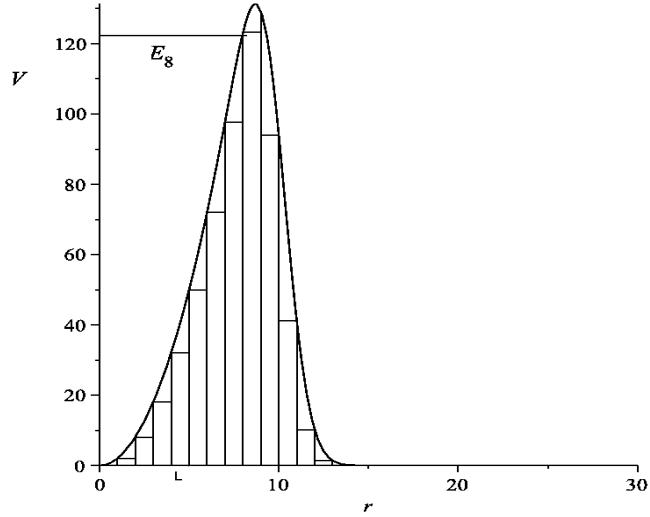}
  \caption{The potential steps approximation.}
\label{potGamov}
\end{figure}
The proper transmission coefficient then becomes:
\begin{eqnarray}\nonumber
T &=& \sum_i 16\frac{E}{V_{0i}}\left ( 1-\frac{E}{V_{0i}} \right ) \exp\left ( -2L \sqrt{\frac{2m}{\hbar^2} (V_{0i}-E)}\right )\\\nonumber
  &=& \sum_i \exp\left ( -2L \sqrt{\frac{2m}{\hbar^2} (V_{0i}-E)} + \ln \left ( 16\frac{E}{V_{0i}}\left ( 1-\frac{E}{V_{0i}} \right ) \right ) \right ).
\end{eqnarray}
To evaluate the transmission coefficient, as presented above, we reconsider the case of $N=11$, $\lambda=0$, $\mu=0.2$ and $l=0$, where we have already calculated the energies and plotted them in figure (\ref{en11lmbd0}). Taking into consideration that $\frac{\hbar^2}{2m}=20.735$\,MeV $\text{fn}^2$ we choose the last energy level $E_8=118.53\,$MeV and with $L=0.96\,$fn the values of $V_i$ are calculated from potential expression (\ref{pot_t_N}). The highest barrier has $129.2776\,$MeV, the transmission coefficient as a function of energy is presented in figure (\ref{tr}),

In conclusion, we obtained the exact bound state solutions of radial Schr\" odinger equation for pseudo-Gaussian potential by using the
wave-function ansatz method. 
We have found that the spectra of a particle trapped into PGO potential allows states with negative energies as well as positive ones.
The resonant states, which are of great interest in the physics of unstable and weakly bound systems like the
halo nuclei, were also discussed and we  have found that they are pretty much stable. However the tunneling coefficient is high compared with literature. 
In spite of infinite power series, which appears in potential terms, we managed to plot the eigen-energies and eigenfunctions for the first levels. 
The results were compared with those obtained by numeric methods.


\begin{thebibliography}{00}
\bibitem{Sch} Schr\" odinger, E. {\em An Undulatory Theory of the Mechanics of Atoms and Molecules}. Phys. Rev. 28 (6): (1926), 1049–1070. 
\bibitem{M} Morse, P. M. {\em Diatomic molecules according to the wave mechanics. II. Vibrational levels}. Phys. Rev. 34. pp. 57–64, (1929).
\bibitem{Eck} Eckart C., {\em The penetration of a potential barrier by electrons}. Phys. Rev. 35(11): 1303-1309; 1930
\bibitem{RM} N. Rosen and Philip M. Morse {\em On the Vibrations of Polyatomic Molecules}. Phys. Rev. 42 pp. 210-217 (1932)
\bibitem{PT} P\" oschl, G.and Teller, E. {\em Bemerkungen zur Quantenmechanik des anharmonischen Oszillators}. Zeitschrift f\" ur Physik 83 (3–4): (1933), 143–151
\bibitem{MR} Millard F. Manning and Nathan Rosen {\em A Potential Function for the Vibrations of Diatomic Molecules}. Phys. Rev. 44 p. 953 (§ 10) (1933)
\bibitem{WS}Woods, R.D.and Saxon, D.S. {\em Diffuse Surface Optical Model for Nucleon-Nuclei Scattering}. Phys. Rev 95 (2): 577–578, (1954).  
\bibitem{Sca} Frederick L. Scarf, {\em New Soluble Energy Band Problem}. Phys. Rev. 112, 1137 (1958) 
\bibitem{Bethe}H. Bethe {\em Zur Theorie der Metalle. I. Eigenwerte und Eigenfunktionen der linearen Atomkette}. Zeitschrift für Physik, 71:205–226 (1931).
\bibitem{Ba}Masao Ogata and Hiroyuki Shiba, {\em Bethe-ansatz wave function, momentum distribution, and spin correlation in the one-dimensional strongly correlated Hubbard model}. Phys. Rev. B 41, 2326 (1990).

\bibitem{Kaushal_other}R.S Kaushal, {\em Quantum mechanics of noncentral harmonic and anharmonic potentials in two-dimensions}. Ann.  Phys. 206, 1, 90–105, (1991)
\bibitem{Kaushal}R.S. Kaushal, {\em An exact solution of the Schrödinger wave equation for a sextic potential}, Phys. Lett. A, 142, 2, 57-58, (1989) 
\bibitem{dong_ans}Shi-Hai Dong {\em On the Solutions of the Schrödinger Equation with some Anharmonic Potentials: Wave Function Ansatz}.  Phys. Scr. 65 289, (2002).
\bibitem{dong_inv}Shi-Hai Dong, 64 273 {\em Schr\" odinger Equation with the Potential $V(r) = Ar^{-4} + Br^{-3} + Cr^{-2} + Dr^{-1} $.},  Phys. Scr. 64 273, (2001) 
\bibitem{Ikhadir} Sameer M. Ikhdair, Ramazan Sever {\em Exact solutions of the radial Schr\" odinger equation for some physical potentials.},  Centr. Eur. J. Phys., 5, 516, 2007, 
\bibitem{GF} Cotaescu I.I, Gravila  P. and Paulescu M, {\em Pseudo-Gaussian oscillators.}  M., Int. J. Mod. Phys. C, 19, 1607-1615 (2008).
\bibitem{F}  F. Iacob, {\em Relativistic pseudo-Gaussian oscillators.}  Phys. Lett. A,  374, 11-12,  1332-1335,  (2010)
\bibitem{Rainville}Rainville E. D. {\em Necessary Conditions for Polynomial Solutions of Certain Riccati Equations.} The American Mathematical Monthly,  43,  8, 473-476, (1936)
\bibitem{dong} Shi-Hai Dong, {\em Exact solutions of the two-dimensional Schr\" odinger equation with certain central potentials}. International Journal of Theoretical Physics 39 (4), 1119-1128 (2000).
\bibitem{Nasser6}Nasser Saad, Richard L Hall and Hakan Ciftci, {\em Sextic anharmonic oscillators and orthogonal polynomials.} J. Phys. A: Math. Gen. 39 8477.(2006)
\bibitem{octic}Agboola D and Zhang Y-Z,  {\em Exact Solutions of the Schr\" odinger Equation with Spherically Symmetric Octic Potential}. Mod. Phys. Lett. A 27, 1250112 (2012) 
\bibitem{dong10} Shi-Hai Dong, {\em Quantum Monodromy in the Spectrum of Schrödinger Equation with a Decatic Potential.} 
International Journal of Theoretical Physics, 41, 1, 89-99, (2002). 
\bibitem{Nasser} David Brandon and Nasser Saad, {\em Exact and approximate solutions to Schr\" odinger's equation with decatic potentials}.  Centr. Eur. J. Phys. 11, 3, 279-290, (2013)
\bibitem{QSS}  A.G Ushveridze {\em Quasi-Exactly Solvable Models in Quantum Mechanics.} Taylor \& Francis Croup NY 10016 (1994).
\end{thebibliography}
\end{document}